\documentclass[usenatbib]{mnras}
\usepackage{epsfig}
\usepackage{amsmath,amssymb}
\graphicspath{{figures/}} 
\usepackage{xcolor}
\usepackage{soul}
\usepackage{aas_macros}  
\usepackage{chngcntr}
\graphicspath{{figures/}}

\newcommand{\Msolar}{\mbox{\,$\rm M_{\odot}$}}        
\newcommand{\Rsolar}{\mbox{\,$\rm R_{\odot}$}}        
\newcommand{\Lsolar}{\mbox{\,$\rm L_{\odot}$}}        

 \newcommand{\teff}{\mbox{\,$T_{\rm eff}$}}      

  \newcommand{\kmsec}{\,\mbox{$\mbox{km}\,\mbox{s}^{-1}$}}    
  \def\simge{\mathrel{\raise1.16pt\hbox{$>$}\kern-7.0pt
    \lower3.06pt\hbox{{$\scriptstyle \sim$}}}}           
  \def\simle{\mathrel{\raise1.16pt\hbox{$<$}\kern-7.0pt
    \lower3.06pt\hbox{{$\scriptstyle \sim$}}}}           
\title[r modes in helium-rich hot subdwarfs]{TESS photometry of helium-rich hot subdwarfs: r modes in BD$+37^{\circ}442$ and BD$+37^{\circ}1977$}
\author[C. S. Jeffery] 
       {C. Simon Jeffery$^1$\thanks{E-mail: simon.jeffery@armagh.ac.uk}\\
$^{1}$Armagh Observatory and Planetarium, College Hill, Armagh BT61 9DG, Northern Ireland }

\date{Accepted .....
      Received ..... ;
      in original form .....}

\pagerange{\pageref{firstpage}--\pageref{lastpage}}
\pubyear{2020}

\begin{document}

\label{firstpage}

\maketitle

\begin{abstract}
{\it TESS} photometry of the extremely helium-rich hot subdwarfs BD$+37^{\circ}442$ and BD$+37^{\circ}1977$ demonstrates multi-periodic low-amplitude variability with principal periods of 0.56 and 1.14\,d, respectively, and with both first and second harmonics present. The lightcurves are not perfectly regular, implying additional periodic and/or non-periodic content. Possible causes are examined, including the binary hypothesis originally introduced to explain X-ray observations, differentially rotating surface inhomogeneities, and pulsations. If the principal photometric periods correspond to the rotation periods, the stars are rotating at approximately 0.7 and 0.3 $\times$ breakup, respectively. Surface Rossby waves (r modes) therefore provide the most likely solution.       
       \end{abstract}


\section{Introduction}
Extreme helium (EHe) stars are rare early-type stars with high luminosity-to-mass ratios ($\log L/M \approx 3 - 4$) and atmospheres virtually devoid of hydrogen \citep{jeffery11a}. Helium-rich hot subdwarfs (HesdO) have atmospheres similarly devoid of hydrogen and  fall into two principal groups: a) stars on or close to the junction of the extended horizontal branch and helium main sequence 
($\log L/M \approx 2 - 3$), and  b) more luminous stars with similarities to the central stars of planetary nebulae ($\log L/M \approx 3 - 4$)  \citep{heber16}. 
Both EHes and luminous HesdOs are considered to be evolving from an asymptotic-giant branch or other helium-shell burning giant configuration to the white dwarf phase \citep{webbink84,saio02,zhang14}.
Connections between these and other classes of hydrogen-deficient and carbon-rich stars have been explored \citep{reindl14}.
 
Critical tools in the exploration of these connections have been studies of stellar surface chemistries \citep{jeffery11a}, duplicity \citep{jeffery87}, 
winds \citep{jeffery10} and pulsations \citep{saio88b,jeffery16a}. 
A large fraction of EHes are photometric variables \citep{jeffery08.ibvs}; two show regular pulsations with periods $\approx 0.1$\,d \citep{landolt75,kilkenny95}.
The remainder have irregular light curves, as clearly demonstrated by recent observations of PV\,Tel and V821\,Cen \citep{jeffery20b}.
Such variability extends from  cool R\,CrB stars with effective temperatures $\approx 7\,000$\,K \citep{alexander72,jones89} through to the hottest {\it bona fide} EHe star V2076\,Oph \citep{lynasgray84,wright06}. 
Learning the boundaries to this type of variability is only limited by the scarcity of EHe stars in the Galaxy. 

Two HesdO stars  BD$+37^{\circ}442$ \citep{rebeirot66} and BD$+37^{\circ}1977$ \citep{,wolff74} are frequently included in lists of EHe stars \citep{drilling86,drilling96}.
Their pulsation characteristics have not yet been realised, but have proved interesting stars for other reasons. 
X-rays have been discovered in both cases \citep{palombara12,palombara15}, prompting speculation that the subdwarf wind collides with a compact companion, as in the hot-subdwarf plus neutron-star binary HD\,49798 \citep{mereghetti16}. 

This paper examines the high-precision photometric behaviour of both stars over a timescale of 30 days, demonstrates milli-magnitude periodic variations, and discusses implications of the latter.  

\begin{figure*}
\epsfig{file=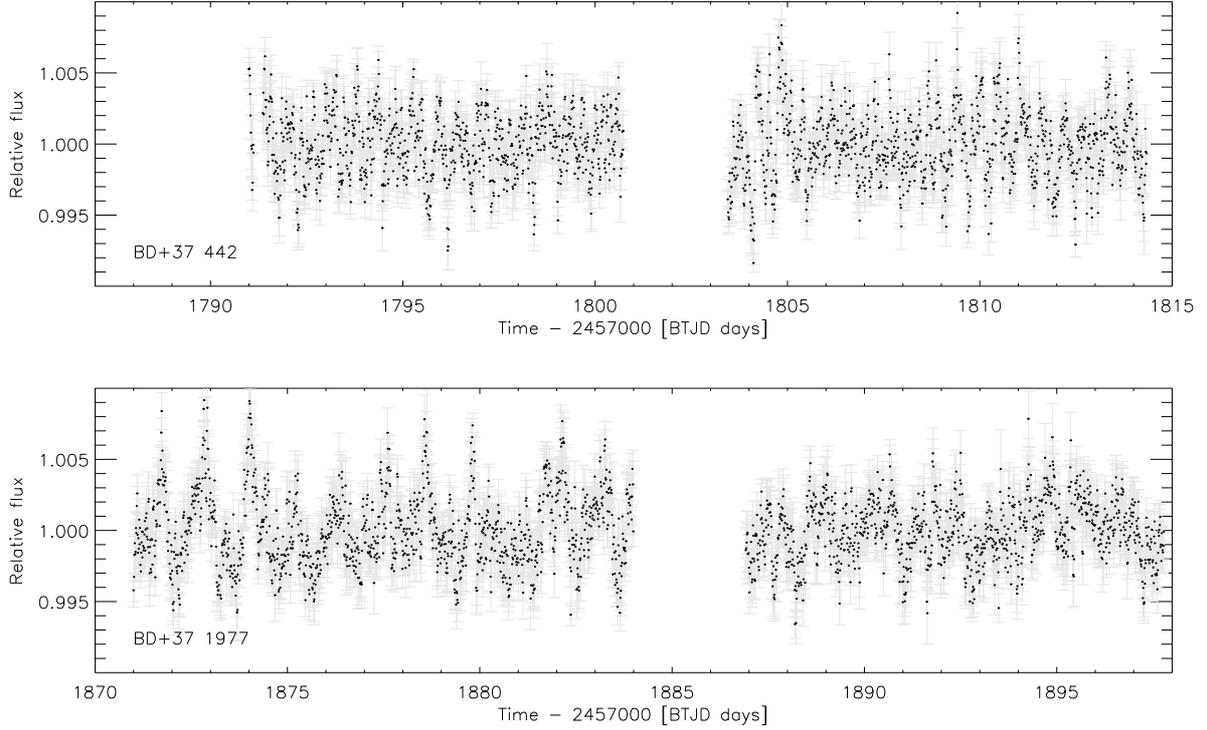,width=160mm,clip=,angle=0}
\caption{{\it TESS} light-curves  spectra for BD$+37^{\circ}442$ and  BD$+37^{\circ}1977$. Errors for each point in the light curves are shown in light grey.} 
\label{f:lc}
\end{figure*}

\section{Observations}

With a primary mission to survey the whole sky for planetary transits, the Transiting Exoplanet Survey Satellite ({\it TESS}) also provides superlative photometry for all manner of variable phenomena. 
Targeted observations were made with {\it TESS} in 120\,s cadence
of BD$+37^{\circ}442$ ($m_V = 9.92$: TIC 067519898) during Sector 18, commencing 2019 November 3, 
and of BD$+37^{\circ}1977$ ($m_V = 10.15$: TIC 372681399) during Sector 21, commencing 2020 January 21.
Standard data products include noise-corrected light curves which are available from the Mikulski Archive for Space Telescopes (MAST). 

\begin{figure*}
\centering
\epsfig{file=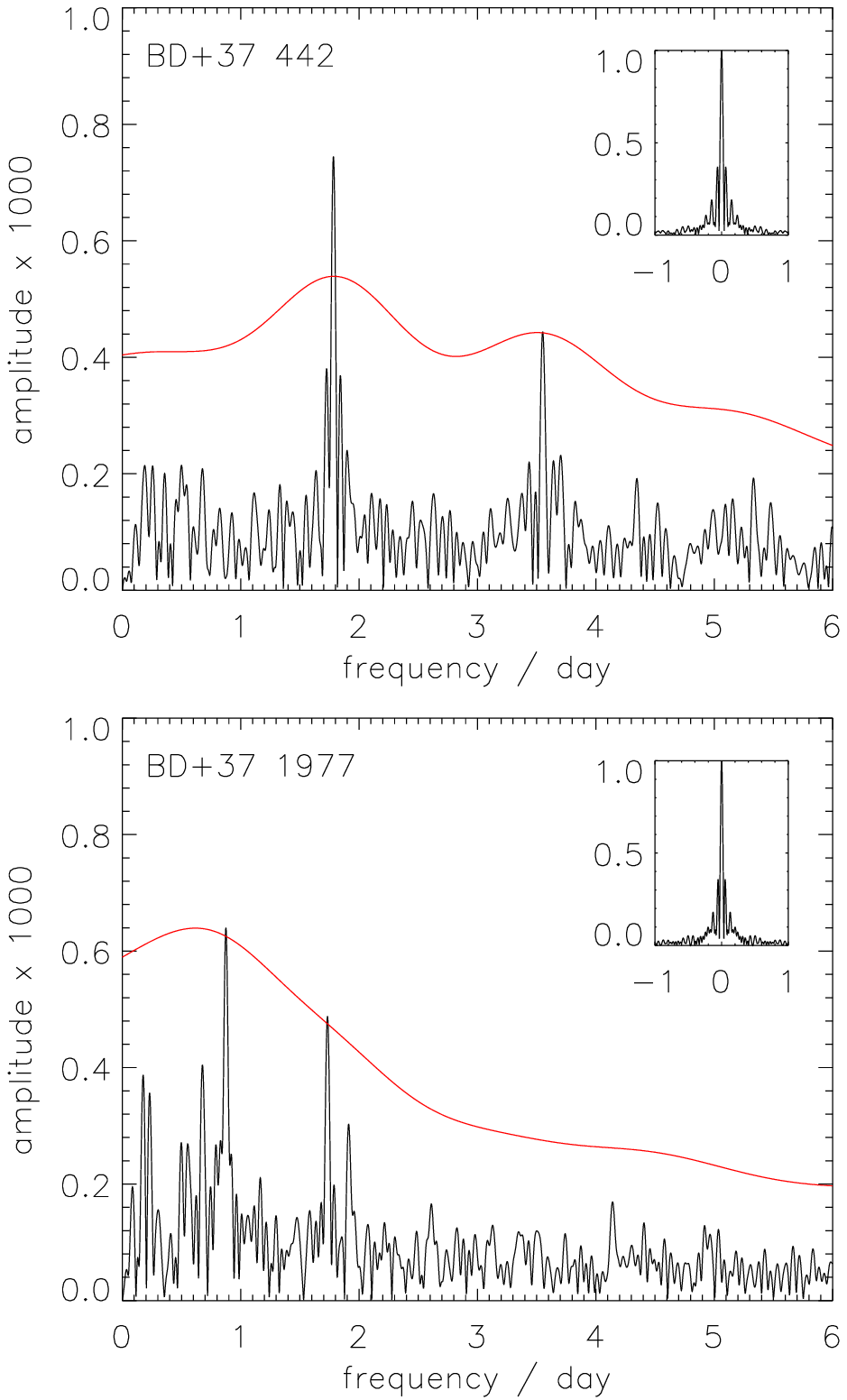,width=70mm,clip=,angle=0}
\epsfig{file=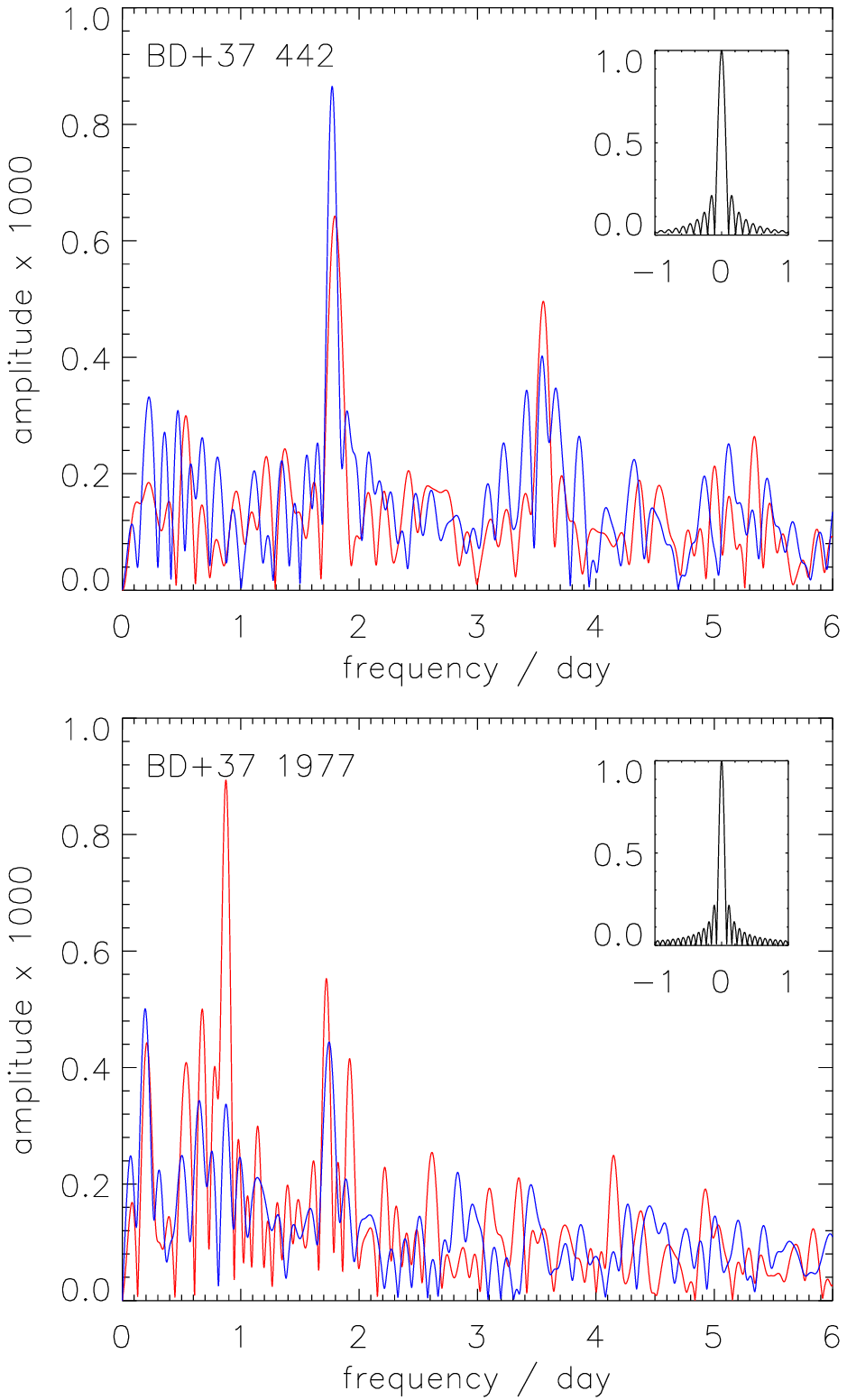,width=70mm,clip=,angle=0}
\caption{Left: frequency-amplitude spectra derived from the {\it TESS} light curves for BD$+37^{\circ}442$ and BD$+37^{\circ}1977$. The spectral window function is inset for both stars. The spectra are essentially flat for frequencies $>2$\,d$^{-1}$. The smooth curve (red online) shows the 4$\sigma$ significance criterion. Right: as above but for the TESS data split into two parts, the transform for each part being shown as red and blue, respectively.
The window function (inset) represents the first (red) segment in each case. 
}  
\label{f:ps}
\end{figure*}

\begin{table*}
    \caption{Principal frequencies, periods and amplitudes in the {\it TESS} light curves.}
    \label{t:freqs}
    \centering
    \begin{tabular}{c ccc ccc}
    \hline
&     \multicolumn{3}{c}{BD$+37^{\circ}442$} & \multicolumn{3}{c}{BD$+37^{\circ}1977$}  \\
$n$ &  $\nu / {\rm d}^{-1}$ &  $P / {\rm d}$ & $ a \% $ & $\nu / {\rm d}^{-1}$ & $P / {\rm d}$ & $ a \% $ \\
\hline
$1$  & $1.78322\pm0.01249$ & 0.5601 & $0.15\pm0.07$ &  $0.87440\pm0.00994$ & 1.1436 & $0.13\pm0.06$ \\
$2$  & $3.55081\pm0.01878$ & 0.2816 & $0.09\pm0.07$ &  $1.73336\pm0.01200$ & 0.5769 & $0.10\pm0.06$ \\
$3$  & $5.33314\pm0.04593$ & 0.1875 & $0.04\pm0.07$ &  $2.60975\pm0.03273$ & 0.3832 & $0.03\pm0.06$ \\
\hline
    \end{tabular}
\end{table*}

\section{Results}

Figure \ref{f:lc} shows the {\it TESS} light-curves of BD$+37^{\circ}442$ and  BD$+37^{\circ}1977$ normalized to their mean flux.  
Both show  variability and appear periodic.
A part of the Fourier transforms of both light curves, in the form of amplitude spectra in the frequency range $0 - 6\,{\rm d^{-1}}$, is shown in Fig. \ref{f:ps}. 
Assuming a $4\sigma$ detection threshold, only two peaks are significant in the amplitude spectra of each star, and all but one of these would normally be marginal. 
However, in both cases, the higher frequency peak is at roughly twice the frequency of the first. \\

\noindent {\bf BD$+37^{\circ}442$} shows variability with a full amplitude of $\approx 1\%$ and two principal peaks in the amplitude spectra corresponding to frequencies of 1.783 and 3.551\,d$^{-1}$ (Table\,\ref{t:freqs}); the ratio of the second frequency to the first is 1.992.  
The amplitude spectrum shows a third peak at a frequency of 5.333\,d$^{-1}$; it is interesting only because it corresponds to 2.991 times the principal frequency.

\noindent {\bf BD$+37^{\circ}1977$} also shows variability with a full amplitude of $\approx 1\%$.
The amplitude spectrum contains more structure, but also shows two principal peaks at frequencies of 0.8744 and 1.7334\,d$^{-1}$ (Table\,\ref{t:freqs}); the ratio is 1.982. 
Again, a third peak a 2.985 times the principal frequency is detectable, but not strong. 
In addition, there are several peaks with frequencies between 0 and 2\,d$^{-1}$
but with amplitudes well below the 4$\sigma$ significance threshold. 

A second significance test is to divide the data into two roughly equal segments and to calculate the amplitude spectrum for each segment (Fig.\ref{f:ps}: right-hand panels). For BD$+37^{\circ}442$, both principal peaks persist although a reduced significance is inevitable. For BD$+37^{\circ}1977$, the strongest peak (0.8744\,d$^{-1}$) is only strong in the first segment; the 1.7334\,d$^{-1}$ is persistent, as is a low frequency peak at 0.1758\,d$^{-1}$. 

BD$+37^{\circ}442$ was reported variable on timescales of minutes with an amplitude of $\approx 3\%$, and months with an amplitude of $\approx 8\%$ \citep{bartolini82}. 
Both BD$+37^{\circ}442$ (= HIP 9221) and BD$+37^{\circ}1977$ (= HIP 46131) were classified `C' (Field H52: `variability not detected') in the Hipparcos catalogue \citep{esa97}. 
The scatter in Hipparcos magnitudes (together with the number of observations) was 0.026 (86) and 0.025 (64) mag. respectively.  X-ray variability with a period of 19 seconds was reported by \citet{palombara12}, but not confirmed by \citet{mereghetti17}.

\begin{table}
    \caption{Fundamental data}
    \label{t:data}
    \centering
    \begin{tabular}{ccc}
    \hline
         & BD$+37^{\circ}442$ & BD$+37^{\circ}1977$  \\
         \hline
    $T_{\rm eff}/{\rm K}$ & $48\,000$ & $48\,000$ \\
    $ \log \dot{M} / \Msolar {\rm yr}^{-1} $&  $-8.5$ & $-8.2$ \\
    & \multicolumn{2}{c}{\citet{jeffery10}} \\[2mm]
    $\log g/ {\rm cm\,s^{-2}}$ & $4.0$ & $4.0$ \\
    $\log L/\Lsolar$ & $4.4$ & $4.4$ \\
      & \multicolumn{2}{c}{\citet{darius79,giddings81} } \\
      & \multicolumn{2}{c}{\citet{bauer95} } \\[2mm]
    $T_{\rm eff}/{\rm K}$ & $56\,300$ & -- \\
    $\log g/ {\rm cm\,s^{-2}}$ & $4.1$ & -- \\
    $ v_{\rm rot,eq} \sin i/{\rm km\,s^{-1}}$ & $60$  & --  \\
    & \multicolumn{2}{c}{\citet{heber14}} \\[2mm]
    $\log L/M / (\Lsolar/\Msolar)^{a}$ & $4.12\pm0.26 $ & $4.12\pm0.26 $ \\[0.5mm]
    $\log \bar{\rho}/\rho_{\odot}\,^{b}$  & $-0.56\pm0.38$ & $-0.56\pm0.38$ \\[0.5mm]
    $P_{\rm F}/{\rm d}\,^{b,c}$ & $0.08\pm0.03$ & $0.08\pm0.03$ \\[0.5mm]
      & \multicolumn{2}{c}{by identity} \\[2mm]
    $m_G$ & 10.13 & 9.95 \\[0.5mm]
    $\pi /{\rm mas}$ & $0.77\pm0.16$ & $0.79\pm0.11$ \\[0.5mm]
         & \multicolumn{2}{c}{\citet{gaia18.dr2}} \\[2mm]
    $d /{\rm kpc}$ & $1.55^{+0.33}_{-0.56}$ & $1.36^{+0.18}_{-0.24}$ \\[0.5mm]
    $ \log L/\Lsolar $ & $4.06^{+0.13}_{-0.33}$ & $3.83^{+0.09}_{-0.15}$ \\[0.5mm]
    $ R/\Rsolar$ & $1.55^{+0.24}_{-0.59}$ & $1.18^{+0.12}_{-0.20}$ \\[0.5mm]
    $ M/\Msolar$ & $0.88^{+0.73}_{-0.77}$ & $0.51^{+0.33}_{-0.26}$ \\[0.5mm]
     & \multicolumn{2}{c}{Martin \& Jeffery (in prep.)} \\[2mm]
    $ P_{\rm rot}/{\rm d}^{d}$  & $0.47 - 0.56$  & $0.96 - 1.15$  \\
    $ v_{\rm rot,eq}/{\rm km\,s^{-1}}^{e}$ & $140 - 170$  & $52 - 62$   \\
    $ P_{\rm min}/{\rm d}^{e}$ & 0.40 & 0.35 \\
    \hline
    \multicolumn{3}{l}{$a$: assuming errors $\pm10\%$ in \teff\ and $\pm0.25$ in $\log g$.}
    \\
    \multicolumn{3}{l}{$b$: assuming mass $M=0.8\Msolar$.}
    \\
    \multicolumn{3}{l}{$c$: assuming $P_{\rm F} \sqrt{\frac{\bar{\rho}}{\rho_{\odot}}} \approx 0.04\,{\rm d}$.}
    \\
    \multicolumn{3}{l}{$d$: from \S\,\ref{s:puls}. }
    \\
    \multicolumn{3}{l}{$e$: note large errors in $R, M$. }
    \end{tabular}
\end{table}


\begin{table}
    \caption{Relative radial velocities for BD$+37^{\circ}1977$ from high-resolution ultraviolet spectroscopy.}
    \label{t:iue}
    \centering
    \begin{tabular}{ccccc}
    Image & Start Time & $\langle \Delta v\rangle$ & $\langle \delta v \rangle$ & $\sigma$   \\
    & & \multicolumn{3}{c}{$  {\rm km\,s^{-1}}$} \\
    swp06766   & 1979-10-05 14:46:59 & $+3.4$ & $\pm$1.7 & 4.3 \\
    swp07248   & 1979-11-28 09:22:12 & $-3.2$ & $\pm$3.5 & 10.6 \\
    tues5247\_5 & 1996-11-28 21:43:05 & $-6.0$ & $\pm$2.6 & 7.0 \\
    tues5247\_7 & 1996-11-29 01:55:13 & $-1.6$ & $\pm$3.1 & 8.9 \\
    tues5247\_8 & 1996-12-02 17:11:03 & $+3.8$ & $\pm$2.4 & 5.8 \\
    tues5247\_9 & 1996-12-02 18:41:19 & $+4.0$ & $\pm$2.6 & 3.5 \\ 
    \end{tabular}
\end{table}

\section{Interpretation}

{\it TESS} observations provide convincing evidence for and characterisation of optical variability in both BD$+37^{\circ}442$ and BD$+37^{\circ}1977$. 
The problem is to identify the cause.
The presence of a harmonic sequence argues for a non-sinusoidal but singly periodic mechanism and suggests either distortion by (or of) a binary companion, by inhomogeneity on the stellar surface (spots) or by a dominant pulsation. 
These are discussed below; additional context is provided by fundamental data for both stars given in Table\,\ref{t:data}.  

\subsection{Double trouble?}


The binary hypothesis resonates with the \citet{palombara12} proposal of a compact companion to explain short-period X-ray flickering in BD$+37^{\circ}442$. 
\citet{heber14} effectively ruled this out from radial-velocity measurements; the absence of X-ray flickering in follow-up observations weakens the proposal further \citep{mereghetti17}. 
\citet{heber14} did find evidence for a high projected rotation velocity $v_{\rm eq} \sin i \approx 60\, {\rm km\,s^{-1}}$ indicating a rotation period $\approx 1$ d. 

We have carried out a similar experiment for BD$+37^{\circ}1977$ using six high-resolution spectrograms obtained with the International Ultraviolet Explorer ({\it IUE}) and the Orbiting Retrievable Far and Extreme Ultraviolet Spectrometers ({\it ORFEUS}) spacecraft, and previously used by \citep{jeffery10} to investigate the stellar wind. 
Spectrograms listed in Table\,\ref{t:iue} were  merged and rectified.
A mean (template) spectrum was constructed from the {\it ORFEUS} data. 
Interstellar lines were identified and removed from the template spectrum where possible. 
Lines formed in the stellar wind \citet{jeffery10} were also avoided. 
Individual spectra were cross-correlated with the template spectrum in four different spectral windows ($\lambda\lambda 1150-1210, 1250-1300, 1300-1400$ and $1150 - 1400$ \AA). 
In this experiment, the cross-correlation function (ccf) generally comprises a narrow spike on top of a broad feature on top of a very broad pedestal \citep[cf. $\upsilon$ Sgr:][]{dudley90}.  
Where present, the narrow spike is due to residual interstellar lines in the template spectrum and provides a check on the reference frame.
The broad pedestal is due to large-scale errors in continuum placement and rectification, which only needs to be crude for this experiment.
The feature between corresponds to the stellar photospheric spectrum, with its center of gravity reflecting the relative velocity relative to that of the template spectrum.  
The positions of all three components were obtained by fitting a multi-component gaussian to the ccf; 
each fit was checked visually and the interstellar and stellar velocities were recorded. 
Results from all four spectral windows were compared for consistency and combined.
The interstellar lines are stable to better than $\pm 2\,{\rm km\,s^{-1}}$.
Table \ref{t:iue} shows the mean photospheric velocity shift of each spectrum relative to the template ($\Delta v$) calculated as the mean obtained from all four spectral windows,  
the mean error $\langle \delta v \rangle$ on $\Delta v$, and the standard deviation $\sigma$. 
The standard deviation and mean error of $\Delta v$ are $4.2\,{\rm km\,s^{-1}}$ and 
 $\pm 2.6\,{\rm km\,s^{-1}}$ respectively. 
These measurements are separated by time intervals of 90 m, 32 h, 5.5 d, 50 d, and 17 y, with no evidence of variability above a few  ${\rm km\,s^{-1}}$.

\begin{figure}
\centering
\epsfig{file=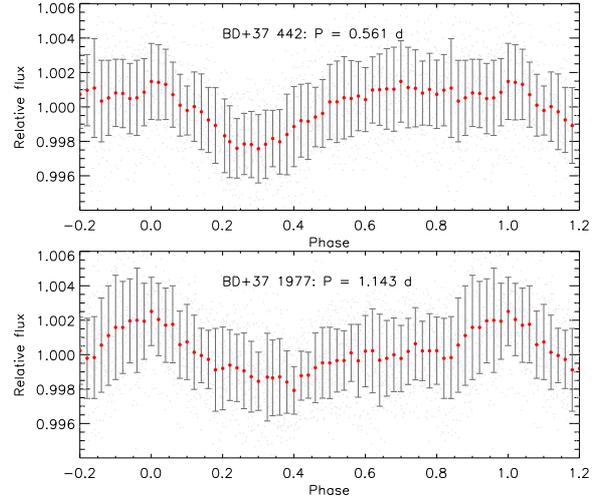,width=82mm,angle=0}
\caption{{\it TESS} light curves for BD$+37^{\circ}442$ and BD$+37^{\circ}1977$ folded on their principal periods. The 120\,s cadence data (faint dots) are combined into phase bins of 0.02 cycles (dark / red dots with standard deviations). 
}  
\label{f:phi}
\end{figure}

The harmonic sequences present in both lightcurves could still argue toward light from a distorted primary or reflected from a secondary in a close binary. 
These would require the harmonics to be {\it exact} and for the lightcurve to be symmetric.
Although $\nu_2$ and $\nu_3$ are harmonics of $\nu_1$ within formal errors (Table\,\ref{t:freqs}) folding the lightcurve on the principal period is unsatisfactory in both cases (Fig.\,\ref{f:phi}). 
Rather than a symmetric or regular light curve with harmonic components, considerable scatter survives after folding, suggesting that other frequencies are important and/or that the harmonics are not {\it exact} integer multiples. 

Hence the evidence for a close binary companion that could give rise to the optical variations is not compelling.

\subsection{All in a spin?} 

Periodic variability can also arise from surface inhomogeneities such as chemical or magnetic spots. 
If distributed over the surface of a uniformly rotating star, the light curve should be dominated by a single period, but with harmonics. 
This does not explain the residual variability about the mean (Fig.\,\ref{f:phi}), or the variation of the amplitude spectrum during an observing run. 
With contraction on a timescale of $10^3 - 10^4$\,y, uniform rotation cannot be assumed \citep{gourgouliatos06}.   
If surface inhomogeneities  migrate in latitude on short time scales, strict and persistent harmonic sequences might not be expected, as different spot patterns acquire different spin periods. 
\citet{howarth19} find that qualitatively similar behaviour in the rapidly rotating bright O4 I(n)fp star $\zeta$ Pup (\citet{howarth14})``would require an ad hoc mechanism of differential rotation coupled to latitudinally mobile hotspots'' and that this argument, also, is not compelling. 

It is interesting to compare photometric periods with the  minimum possible stellar rotation period for a positive equatorial effective gravity $P_{\rm min} = 3\pi\sqrt{R_{\rm eq}/g_{\rm p}} = 0.39$\,d for both stars (Table\,\ref{t:data}). 
Here we assume the equatorial radius $R_{\rm eq} = \sqrt{1.5}R$ and the polar gravity $g_{\rm p}=g$.
In other words, if $P_1 \approx P_{\rm rot}$, both stars are rotating rapidly.
This is relevant in \S\,\ref{s:puls}, but also places a strong constraint on
the surface gravities  $\log g/{\rm cm\,s^{-2}} \geq 3.8$ and 3.4 for 
BD$+37^{\circ}442$ and BD$+37^{\circ}1977$ respectively. 
Published measurements comfortably exceed these limits (Table\,\ref{t:data}). 

\subsection{Riding the wave?} 
\label{s:puls}

For a star of given dimensions, the mean density provides an approximate value for the period of the fundamental radial mode \citep[$P_{\rm F}$:][]{eddington18}. 
Corresponding values for both programme stars are $\approx0.08$\,d, subject to assumptions
given in Table\,\ref{t:data}. 
The periods corresponding to principal frequencies observed in both light curves (Table\,\ref{t:freqs}) are approximately 18 and 9 times longer than $P_{\rm F}$, respectively. 
Conventional non-radial pulsations at these frequencies would have to be g modes. However, the absence of equal period intervals, the period length relative to $P_{\rm F}$, and evidence for  harmonic sequences ($\nu_1, 2\nu_1, 3\nu_1$) for both stars argue against g modes. 

An alternative is that Rossby waves, otherwise known as r modes, account for the observations \citep{saio82,townsend03b}. 
These are retrograde surface waves excited in gaseous surface layers of rotating objects (including both stars and planets) and are increasingly being discovered in rapidly-rotating early-type stars \citep{vanreeth16,saio18}.
Depending on the inclination, r modes are predicted to be visible in frequency groups close to harmonics of the rotation frequency, with the harmonic corresponding to the azimuthal wave number of the non-radial mode $m$. 

Considering the most visible modes symmetric about the equator (i.e. $k=-2$, where $k$ is a parameter used to order eigenfunctions of the Laplace tidal equation \citep{lee97,townsend03b}), 
the expected observational frequency range of r modes ($\nu_r$) is given by
\[m f_{\rm rot}[1 - 2/[(m+1)(m+2)] ] <  \nu_r < m f_{\rm rot},\]
where the sectoral wavenumber $m > 0$ for retrograde modes such as r modes \citep{saio18b} 
and $f_{\rm rot}$ is the rotation frequency. 
If the two principal frequencies observed in each star correspond to r modes with $m=1$ and 2, rotational frequencies are then given by
\[ \nu_{r,m=1} < f_{\rm rot} < \frac{3}{2}\nu_{r,m=1}\]
and
\[ \frac{1}{2} \nu_{r,m=2}  < f_{\rm rot} < \frac{3}{5}\nu_{r,m=2},\]
and are consistent with r modes if the rotation periods $P_{\rm rot}$ of the two stars lie in the ranges 0.47 -- 0.56 d (BD$+37^{\circ}442$) and 0.96 -- 1.15 d (BD$+37^{\circ}1977$).
For BD$+37^{\circ}442$, this implies an equatorial velocity in the range 140 -- 170 \kmsec, consistent with the \citet{heber14} measurement with inclination $i\approx 30^{\circ}$.    


\section{Conclusion}

We have presented {\it TESS} light curves for two extreme helium hot subdwarfs. 
Both are interesting as sources of X-rays generated in a stellar wind, and as putative descendants of other extremely helium-rich stars. 
The {\it TESS} data provide the first unequivocal evidence for light variability in either star, variability which is clearly periodic. 
Discerning the period driver is more difficult, but it is noted that the principal photometric periods are relatively large fractions of the theoretical minimum rotation periods for breakup.  
An absence of variations in radial velocity argues against either star being a member of a short-period binary. 
Apparent harmonics in the frequency spectra are not exact, but this could be due to the limited length of the {\it TESS} light curves. 
Folding the light curves on the principle periods provides scant help; the average shape is asymmetric and the residuals remain substantial.
The most satisfactory explanation is that the principal periods and their harmonics correspond to Rossby waves excited on the surface of two quite rapidly rotating hot subdwarfs. 
{\it TESS} observations of other hot subdwarfs should be inspected for harmonic sequences of low-frequency signals; corresponding evidence of a high rotation velocity would strongly suggest an r mode origin. 

\section*{Acknowledgments}
This paper includes data collected with the {\it TESS} mission, obtained from the MAST data archive at the Space Telescope Science Institute (STScI). 
Both of the stars studied were placed on the 2 min cadence roster by Working Group 8 (Evolved Compact Stars) of the {\it TESS} Asteroseismic Science Operations Center. 
Funding for the {\it TESS} mission is provided by the NASA Explorer Program. STScI is operated by the Association of Universities for Research in Astronomy, Inc., under NASA contract NAS 5–26555. 

\bibliographystyle{mnras}
\bibliography{ehe}
\label{lastpage}
\end{document}